# The Application of Techniques Derived from Artificial Intelligence to the Prediction of the Solvency of Bank Customers: Case of the Application of the CART type Decision Tree (DT)

Karim Amzile and Rajaa Amzile

Faculty of Law, Economics and Social Sciences Agdal.
Mohammed V University of Rabat, Morocco

## ABSTRACT

*In this study we applied the CART-type Decision Tree (DT-CART) method derived from artificial intelligence technique to the prediction of the solvency of bank customers, for this we used historical data of bank customers. However we have adopted the process of Data Mining techniques, for this purpose we started with a data preprocessing in which we clean the data and we deleted all rows with outliers or missing values as well as rows with empty columns, then we fixed the variable to be explained (dependent or Target) and we also thought to eliminate all explanatory (independent) variables that are not significant using univariate analysis as well as the correlation matrix, then we applied our CART decision tree method using the SPSS tool.*

*After completing our process of building our model (AD-CART), we started the process of evaluating and testing the performance of our model, by which we found that the accuracy and precision of our model is 71%, so we calculated the error ratios, and we found that the error rate equal to 29%, this allowed us to conclude that our model at a fairly good level in terms of precision, predictability and very precisely in predicting the solvency of our banking customers.*

## KEYWORDS

*Data Mining, Credit Risk, Bank, Decision Tree, Artificial Intelligence, Credit, risk.*

## 1. INTRODUCTION

A Decision Tree (DT) is one of the methods of decision support, originating from Data Mining techniques. Its structure is based on a set of choices in the graphic form of a tree. The various possible decisions are at the end of the branches, and are reached according to decisions taken at each stage of the learning process. DT is one of the methods of supervised learning techniques, it allows to model a hierarchy of dichotomous tests and it also allows to predict the class of individuals in the case of a variable to be explained binary, such as the case of the solvency of bank customers.

A Decision Tree (DT) is one of the methods of decision support, originating from Data Mining techniques. Its structure is based on a set of choices in the graphic form of a tree. The various possible decisions are at the end of the branches, and are reached according to decisions taken at each stage of the learning process. DT is one of the methods of supervised learning techniques, it allows to model a hierarchy of dichotomous tests and it also allows to predict the class of individuals in the case of a variable to be explained binary, such as the case of the solvency of bank customers.

Each DT technique uses its own fractionation algorithms and fractionation measurements. Using well-known DT algorithms, the ID3, C5.0 and C&RT (CART) methods use impurity measurements to select the fractionation attribute and the fractionation value/s. While ID3 (Han et al 2000) uses information gain (IG[1]), C5.0 uses the information gain ratio (IGR[2]), and CART [3](Koh et al 2004) uses the Gini Index (GI[4]) for impurity measurements. For CHAID uses the KHI-2 or the F statistic to choose the division variable (Koh et al 2004).

## 2. LITERATURE REVIEW

In recent years, several sectors have adopted DM techniques to analyze dependent variables (Cheng and Leu 2011, Cheng et al 2015; Lee et al 2016). Although DM techniques have been applied to the analysis of banking data with the aim of predicting the behavior of decisive or explaining variables. Several methods and algorithms, such as neural networks, SVM and DT, have been used in DM to perform data classification and predictive modelling (Ryua et al 2007; Kim 2008).

DM techniques have six functions: association, grouping, classification, prediction, estimation and sequencing. Classification is a supervised analysis function and the most commonly used risk management function (Efraim et al, 2010). Several methods from this category of methods, among these analytical methods for classification and prediction.

DT has a structure that is easy to understand and explain. Thus, a tree diagram is often used to solve a series of classification and decision-making problems to classify many attributes and make predictions; for example, such diagrams have been applied in medical diagnosis and disease classification studies (Khan et al, 2009; Ture, Tokatli, & Kurt, 2009). A DT can also be used for classification studies in engineering and management (Syachrani & Chung, 2013; Alipour et al., 2017).

## 3. THE COMPONENTS OF A DECISION TREE

A DT is a hierarchical, triangular staircase construction consisting of several elements:

- Nodes: There are two categories of nodes:

    ✓ Root node (NR): the first node in the tree, which concerns the first test.
    ✓ Intermediate nodes (NI): nodes from the lower levels provide tests to better partition the individuals in our sample, for these nodes, they receive arrows in inputs and send arrows in outputs.

---

[1]Information Gain (GI) measures the amount of information that an explanatory variable gives us about the variable to explain. It tells us how important a descriptor is. Information Gain (IG) is used to decide the order of independent variables in the nodes of a decision tree.
[2]The information gain ratio is a relationship between the information gain and the intrinsic information.
[3]Classification and Régression Trees (Classification et arbre de régression)
[4]Is a statistical measure of the distribution of a variable in a population.

• Arrows: They indicate the direction of the path to be followed until the final decision. In our case we use dichotomous trees, that is, from each node, we will have two outgoing arrows (True/ False); one indicates the decision to be made or the next test if the condition in the node is true, and the other if not.

• Leaves: are nodes of the last levels of the tree, they contain the decision or class predicted for the individuals. For this kind of knot, there are only inward arrows for these leaves.

## 4. THE CONSTRUCTION OF AN AD OF TYPE CART:

A Decision Tree of type CART proposed by (Beriman et al, 1984) is a binary tree that helps in the decision of a dichotomous variable (Y) for an individual whose explanatory variables are known. Its construction is based on recursive partitioning of individuals using learning data. This partitioning is done by a succession of severed nodes using the Gini index.

To this end, to build a CART tree, we must answer these questions:

A. The determination of a criterion to select the best division among all those eligible for the different variables.
B. A rule to decide that a node is terminal.
C. The assignment of each sheet to one of the classes (for the variable to be explained qualitative) or to a value of the variable to be explained (for the quantitative case).

To answer the above questions the following steps are followed:

A. The division criterion is based on the definition of a heterogeneity function. The objective is to subdivide individuals into two segments which are the most homogeneous in the sense of the variable to be explained. The heterogeneity of a node is measured by a function that returns 2 types of results:

   a. Zero if, and only if, the node is homogeneous, that is, the individuals belong to the same modality of the variable to be explained.
   b. Maximum when Y values are equiprobable (Over-learning) or highly dispersed (Under-learning).

For our case, to choose the root variable or the segmentation variable, that is relative to the top of our decision tree, for this purpose we will use the index of Gini[5], this index adopted by the algorithm CART, measures the frequency for an individual to be misclassified by the node test. This will allow us to choose the variable that achieves the best separation of individuals, we will use a segmentation criterion that we calculate for the different learning variables.

To calculate the value of the Gini-index the following formula is used:

*Équation 1*

$$GI(D) = \sum_{i=1}^{k} p_i(1 - p_i)$$

With:

---

[5]The Gini Index is a variant value from 0 to 1, where 0 means perfect equality and 1 means perfect inequality

$D$: all data used
$k$: The number of modalities of the dependent variable (Solvable or Non-Solvable)
$p_i$: Represents the probability of modality i (creditworthy or non-financial) in the sheet after segmentation.

For our case we have 2 modalities so we can write:

*Équation 2*
$$GI(D) = p_1(1 - p_1) + p_2(1 - p_2)$$
$$GI(D) = p_1 - p_1^2 + p_2 - p_2^2$$

We have:

$$p_1 + p_2 = 1$$

Therefore:

*Équation 3*
$$GI(D) = 1 - p_1^2 - p_2^2 = 1 - \sum_{i=1}^{k} p_i^2$$

B. To decide in which node we will stop, although the growth of the tree must stop at a given node, which thus becomes terminal or leaf. While we will stop the segmentation of the node, when it is homogeneous or when there is no longer an admissible partition or if the number of observations it contains is less than a threshold value (between 1 and 5).

C. The assignment of sheets to one of the classes depends on the type of variable to be explained (Y),

• For quantitative case Y, each sheet has a value that is the average of the observations associated with that sheet.
• In the qualitative case, each sheet or terminal node is assigned to a class (Solvable / Non-Solvable) of Y by considering one of the following conditional modes:

• The class best represented in the node;
• The most likely posterior class in the Bayesian sense;

## 5. THE PREDICTION OF THE SOLVENCY OF A BANK'S CUSTOMERS USING THE CART-DT:

### 5.1. Pre-Data Processing

**a. variable definitions**

For our study, we used data from a bank's customers, our database consists of 3988 rows and 14 columns, among the variables used we have 13 explanatory variables and a single variable to explain which represents the solvency of customers.

To ensure a better modelling we thought about the codification of nominal qualitative variables. The table below illustrates the nature of each variable used and its values after coding:

Table 1: Coding of variables used

| | |
|---|---|
| | **NAME_CONTRACT_TYPE(Boolean)** |
| **Cash loans** | 1 |
| **Revolving loans** | 0 |
| | **CODE_GENDER (Boolean)** |
| **F** | 1 |
| **M** | 0 |
| | **FLAG_OWN_CAR (Boolean)** |
| **Y** | 1 |
| **N** | 0 |
| | **NAME_INCOME_TYPE (Qualitative)** |
| **State servant** | 1 |
| **Working** | 2 |
| **Commercial associate** | 3 |
| **Pensioner** | 4 |
| | **NAME_FAMILY_STATUS (Qualitative)** |
| **Married** | 1 |
| **Single / not married** | 2 |
| **Civil marriage** | 3 |
| **Separated** | 4 |
| **Widow** | 5 |
| | **NAME_HOUSING_TYPE (Qualitative)** |
| **House / apartment** | 1 |
| **With parents** | 2 |
| **Municipal apartment** | 3 |
| **Office apartment** | 4 |
| **Co-op apartment** | 5 |
| **Rented apartment** | 6 |
| | **NAME_EDUCATION_TYPE (Qualitative)** |
| **Higher education** | 1 |
| **Incomplete higher** | 2 |
| **Secondary / secondary special** | 3 |
| **Lower secondary** | 4 |

**b. univariate analysis**

Based on the result of the univariate analysis on SPSS, we obtained the following table 2:

Table 2 : univariate analysis

| Vi | L'intitulé de la variable | B | E.S. | Wald | ddl | Sig |
|---|---|---|---|---|---|---|
| $V_2$ | NAME_CONTRACT_TYPE | ,429 | ,132 | 10,522 | 1 | ,001 |
| $V_3$ | CODE_GENDER | -1,604 | ,165 | 94,361 | 1 | ,000 |
| $V_4$ | FLAG_OW0_CAR | -,424 | ,085 | 24,811 | 1 | ,000 |
| $V_5$ | CNT_CHILDREN | ,341 | ,126 | 7,298 | 1 | ,007 |
| $V_6$ | AMT_INCOME_TOTAL | ,000 | ,000 | 39,407 | 1 | ,000 |
| $V_7$ | AMT_CREDIT | ,000 | ,000 | 34,029 | 1 | ,000 |
| $V_8$ | **AMT_ANNUITY** | ,000 | ,000 | ,435 | 1 | ,510 |
| $V_9$ | AMT_GOODS_PRICE | ,000 | ,000 | 44,091 | 1 | ,000 |
| $V_{10}$ | NAME_INCOME_TYPE | -,375 | ,063 | 35,415 | 1 | ,000 |
| $V_{11}$ | NAME_EDUCATION_TYPE | ,746 | ,038 | 375,557 | 1 | ,000 |
| $V_{12}$ | **NAME_FAMILY_STATUS** | -,017 | ,043 | ,151 | 1 | ,697 |
| $V_{13}$ | NAME_HOUSING_TYPE | ,108 | ,044 | 6,167 | 1 | ,013 |
| $V_{14}$ | CNT_FAM_MEMBERS | -,263 | ,111 | 5,650 | 1 | ,017 |
|  | **Constante** | 1,497 | ,354 | 17,887 | 1 | ,000 |

Based on the univariate analysis, it can be concluded that the variables V_8 (AMT_ANNUITY) and V_12 (NAME_FAMILY_STATUS) are not significant because of their meaning values that exceed 5%, however they must be eliminated in order to build a powerful model in terms of significance by keeping only the variables that have a significant probability.

However, we can also check for qualitative explanatory variables, if there are correlations between dependent variables, to this effect we draw the following correlation matrix (Table: 3):

Table 3 : correlations matrix

| | V2 | V3 | V4 | V5 | V6 | V7 | V9 | V10 | V11 | V13 | V14 |
|---|---|---|---|---|---|---|---|---|---|---|---|
| *correlations* | | | | | | | | | | | |
| V2 | 1 | -,023 | -,034* | -,004 | ,000 | ,227 | ,188 | -,027 | ,053 | -,003 | ,007 |
| V3 | -,023 | 1 | -,093 | ,059 | -,026 | ,063 | ,062 | ,001 | -,049 | -,037 | ,079 |
| V4 | -,034* | -,093 | 1 | ,083 | ,179 | ,108 | ,116 | ,024 | -,146 | -,053 | ,109 |
| V5 | -,004 | ,05 | ,083 | 1 | -,056 | -,033 | -,042 | -,013 | -,014 | ,014 | ,888 |
| V6 | ,000 | -,026 | ,179 | -,056 | 1 | ,379 | ,388 | ,196 | -,273 | -,058 | -,067 |
| V7 | ,227 | ,063 | ,108 | -,033 | ,379 | 1 | ,986** | ,106 | -,197 | -,072 | ,012 |
| V9 | ,188** | ,062** | ,116** | -,042** | ,388 | ,986** | 1 | ,108 | -,215 | -,073 | ,005 |
| V10 | -,027 | ,001 | ,024 | -,013 | ,196 | ,106 | ,108 | 1 | -,104 | -,005 | -,028 |
| V11 | ,053 | -,049 | -,146 | -,014 | -,273 | -,197 | -,215 | -,104 | 1 | ,074 | -,010 |
| V13 | -,003 | -,037 | -,053 | ,014 | -,058 | -,072 | -,073 | -,005 | ,074 | 1 | -,015 |
| V14 | ,007 | ,079 | ,109 | ,888 | -,067 | ,012 | ,005 | -,028 | -,010 | -,015 | 1 |

According to the above matrix, we can notice that there is a strong correlation between V_9 and V_7 so between V_14 and V_5, for this purpose we will proceed to eliminate a single variable in each strongly correlated couple, to ensure better predictability of the model. For this we will eliminate V_9 and V _14.

So, we will keep only the following variables:

| $V_i$ | L'intitulé de la variable |
|---|---|
| 1 | NAME_CONTRACT_TYPE |
| 2 | CODE_GENDER |
| 3 | FLAG_OW0_CAR |
| 4 | CNT_CHILDREN |
| 5 | AMT_INCOME_TOTAL |
| 6 | AMT_CREDIT |
| 7 | NAME_INCOME_TYPE |
| 8 | NAME_EDUCATION_TYPE |
| 9 | NAME_HOUSING_TYPE |

## 6. THE GRAPHICAL REPRESENTATION OF THE AD CART

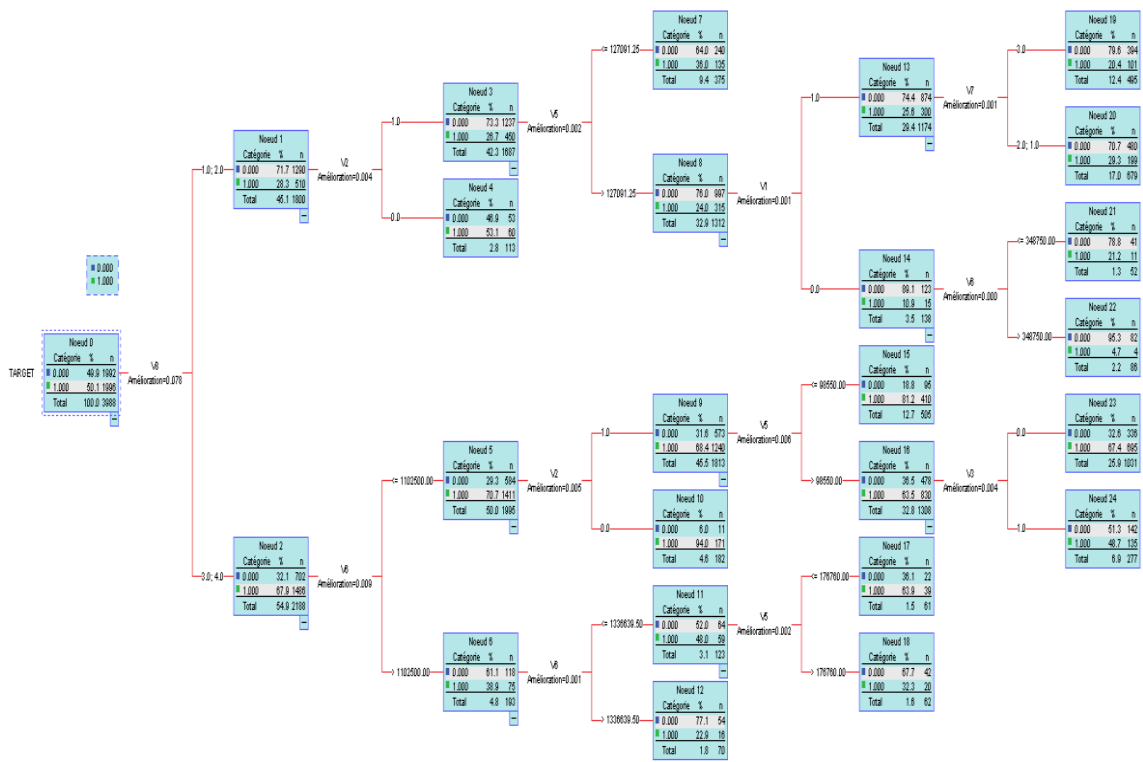

Figure 1 : DT- CART (SPSS / Source : Author)

## 7. ASSESSMENT OF OUR MODEL

### 7.1. Confusion Matrix

| Observations | Classification | | |
|---|---|---|---|
| | Prediction | | |
| | 0 | 1 | Pourcentage correct |
| 0 | 1475 | 517 | 74,0% |
| 1 | 621 | 1375 | 68,9% |
| Pourcentage global | 52,6% | 47,4% | 71,5% |
| Method : CART  Variable dependent : TARGET | | | |

True Positives (VP): 1 correctly predicted VP = 1375
True Negatives (VN): 0 correctly predicted VN = 1475
False Negatives (FN): 0 incorrectly predicted FN = 517
False Positives (FP): 1 incorrectly predicted FP = 621
Sum = 3988

### 7.2. Measurement of error rates

We will use three error ratios:

$$e_1 = \frac{number\ of\ observations(1)\ classified(0)}{number\ of\ obs(1)}$$

$$e_2 = \frac{number\ of\ obs(0)\ classified\ (1)}{number\ of\ obs(1)}$$

$$e_3 = \frac{number\ of\ obs(1) classified(0) + number\ of\ obs(0)\ classified\ (1)}{number\ of\ obs}$$

To calculate these 3 types of errors we must build our confusion matrix:

So, we can now calculate the errors:

$$e_1^{cart} = \frac{621}{1996} \quad e_2^{cart} = \frac{517}{1992} \quad e_3^{cart} = \frac{1138}{3988}$$

| | $e_1$ | $e_2$ | $e_3$ |
|---|---|---|---|
| **AD-Cart** | 0,31112224 | 0,25953815 | 0,285356 |

So according to the above matrix we can calculate the following metrics:

$$Accuracy = \frac{VP + VN}{Somme} = \frac{1375 + 1475}{3988} = 0.71$$

$$Sensibilité = \frac{VP}{VP + FP} = \frac{1375}{1375 + 621} = 0.69$$

$$Spécificité = \frac{VN}{VP + FP} = \frac{1475}{1375 + 621} = 0.74$$

### 7.3. the ROC curves

| Area Under the curve | | | | |
|---|---|---|---|---|
| Zone | Erreur Std | Signification | Confidence interval 95% asymptotique | |
| | | | Lower Bound | Upper bound |
| AUC = 0,715 | 0,008 | 0,000 << 0.05 | 0,698 | 0,731 |

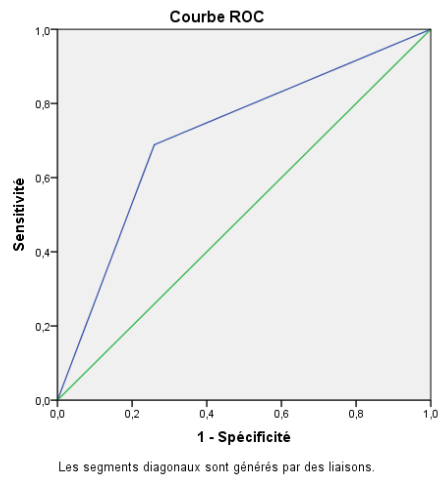

Figure 2 : ROC Curve

## 8. OUTCOME AND DISCUSSION

According to the previous illustrations of the evaluation results of our CART-type DT model, we were able to obtain a model with a precision of 0.71 which is quite good and we also noticed this precision in the graphical representation of the ROC curve, of which we obtained an area under the curve equal to $AUC = 0.715$. However, our model according to the values of $specificity = 0.74$ as well as $Sensitivity = 0.69$ reflects that our model is better performing in the classification of non-bankable clients than creditworthy clients.

## 9. CONCLUSION

After applying DT-CART in our study, we obtained satisfactory results; in terms of precision, as a result our model achieved an accuracy of 71%, therefore the error rate e_3 does not exceed 29%, and the ROC curve reflects a fairly good percentage of precision. Therefore, we can say that we have been able to build a model of new machine learning methods, with a fairly good performance and predictability level.

By way of conclusion, we can say that the techniques of Data Mining derived from artificial intelligence, can be used to solve problems of management of banking risks, to this end, we recommend exploiting this type of technique in the various types of financial risks, in order to reveal even more the opportunity offered by these methods to financial institutions and specifically to banks.

## AUTHORS

**Karim AMZILE** (PhD, Student)
Faculty of Law, Economics and Social Sciences Agdal.
Mohammed V University of Rabat, Morocco

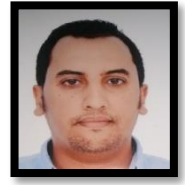

**Rajaa AMZILE** (PhD, Professor)
Faculty of Law, Economics and Social Sciences Agdal.
Mohammed V University of Rabat, Morocco

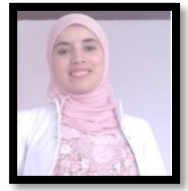